\documentclass[12pt]{iopart}

\usepackage{graphicx} 
\usepackage{url} 
\usepackage{doi}

\newcommand{\be}{\begin{equation}}
\newcommand{\ee}{\end{equation}}

\begin{document}


\title{Black holes, Achilles and the original tortoise coordinate}

\author{Markus P\"ossel}
\address{Haus der Astronomie and Max Planck Institute for Astronomy, K\"onigstuhl 17, 69117 Heidelberg, Germany}
\ead{poessel@hda-hd.de}
\vspace{10pt}
\begin{indented}
\item[]16 July 2024
\end{indented}

\date{\today}

\begin{abstract}
The ``frozen star'' picture of black holes, based on the fact that an outside observer will never see a collapsing sphere shrink sufficiently to form a black hole horizon, was a historical obstacle to the understanding of black holes, and continues to be a stumbling point for students trying to understand horizons. Promoting the discrete steps in Zeno's original story of Achilles and the tortoise to a ``Zeno time coordinate'' provides a quantitative toy model that allows students to understand why the ``frozen star'' phenomenon  does not mean that objects cannot fall into a black hole. The toy model can be used for teaching about this particular feature of black holes in an introductory setting that does not introduce the Schwarzschild metric and its tortoise coordinates.\footnote{This is the version of the article as submitted to the {\em European Journal of Physics}, before peer review. The version of record can be found at Eur.\ J.\ Phys.\ under doi \href{https://doi.org/10.1088/1361-6404/ad82b9}{10.1088/1361-6404/ad82b9}}
\end{abstract}

\maketitle 

Zeno's story about Achilles and the tortoise, which has come down to us via Aristotle (tortoise not included) and Simplicius of Cilicia (with tortoise), is well-known.\cite{StanfordEnc} Zeno argues that Achilles can never catch up with a tortoise, as long as the tortoise has a head start. In modern terms: Assume that at a specific time, the tortoise has reached the location $x_0$. As Achilles catches up, he will also pass the location $x_0$. But by that time, the tortoise has moved a bit further, say to location $x_1$. And once Achilles has reached location $x_1$, the tortoise has moved on a bit further still, to location $x_2$. Whenever Achilles reaches one of the tortoise's previous locations, the tortoise will already have moved on, so Achilles can never catch up with the tortoise.

Achilles and the tortoise are an interesting pre-calculus example of conceptual problems with infinitesimally small divisions, mirrored by students' problems in understanding infinite processes.\cite{deLuca2020a} In physics teaching, a pertinent problem is that of a bouncing ball, which looses height with each bounce.\cite{deLuca2020b} Just like the bouncing ball, the original Achilles-tortoise set-up is an example of a geometric progression.\cite{Stauffer1988}

In the general relativity literature, the problem lives on in the ``tortoise coordinate,'' first introduced by Wheeler,\cite{Wheeler1955} popularized by Regge and Wheeler,\cite{Regge1957} and later communicated to generations of students in the seminal Misner-Thorne-Wheeler text book.\cite{MTW1973}

This coordinate, in turn, is linked to an interesting pedagogical problem in teaching about black holes: Writing down the metric of the simplest (spherically symmetric) black hole in the original form found by Karl Schwarzschild, and calculating the time it takes a test particle to fall into the black hole, one finds that, in the coordinates used by Schwarzschild, it takes the particle an infinitely long (coordinate) time to reach the black hole horizon. The same is true for a sphere that is collapsing to form a black hole: For an outside observer, that sphere comes arbitrarily close to forming a horizon, but does not form a black hole in finite (coordinate) time.

This unusual feature has played an important role in delaying physicists' understanding of the concept of black holes. The ``golden age'' of black hole physics could only begin once it had been realised that the seeming impossibility of falling into a black hole is merely a coordinate effect, and that black holes are more than just ``frozen stars'' hovering forever at a limiting surface.\cite{Israel1987,Thorne1994} Regge-Wheeler tortoise coordinates and related coordinates (notably Eddington-Finkelstein) provide a description that makes it possible to readily follow an infalling object beyond the event horizon. 

The ``frozen star'' effect is directly related to the connection between the Schwarzschild coordinates and the perspective of an observer at infinity, whose coordinates are based on receiving light and, following a recipe similar to that of special relativity, reconstructing when that light must have left its source. Since, by definition, light from inside the event horizon can never reach an observer at infinity, this kind of coordinate definition naturally fails at the horizon. Indeed, we can never see something fall into a black hole --- light conveying information to us will be redshifted to arbitrarily large wavelengths as the light source approaches the horizon, and will take arbitrarily long to reach an outside observer.\cite{EBH} Prose descriptions of this phenomenon appear in a number of popular-science texts on black holes.\cite{Greenstein1983,Thorne1994,Bartusiak2015,Falcke2021}

While the link between the ``frozen star'' phenomenon and the story of Achilles and the tortoise via the Regge-Wheeler tortoise coordinates is a staple of the introductory literature, such coordinate descriptions are too advanced for conceptual introductions to general relativity and black holes that make do without introducing the concept of a spacetime metric, e.g. in the context of an astronomy and astrophysics 101 course or an overview of modern physics for non-majors. 

In this note, I would like to provide the connection between the (advanced) geometric tortoise coordinates and the analysis of Zeno's problem in the fundamental mathematics literature that I have found helpful in teaching non-physics-majors about this aspect of black holes. The courses in question cover their subject without introducing the advanced mathematical formalities (no metric, no tensors), but they do make use of high-school level mathematics, and they introduce a number of relevant simplified calculations that go beyond a purely popular-science account. What follows is an example of such a simplified calculation, aimed at illustrating just what was happening when physicists regarded black holes as ``frozen stars.'' 

We treat the ``observational steps'' of Zeno's description as a discrete time coordinate $T$, and generalise it to a continuous ``Zeno time coordinate,'' as in \cite{Feng2023}. We also describe the situation with the usual Cartesian coordinate $x$, which we take to be defined parallel to the race track, and a time coordinate $t$ that is classical (clock) time. In those coordinates, we assume the speed of Achilles to be a constant $v$, and that of the tortoise a constant $w<v$.

At the time $t=0$, let Achilles be at $x=0$, and the tortoise is at $x=x_0$. We choose this as the zero point for our new time coordinate as well, $T=0$. The next step considered by Zeno, which we identify with $T=1$, is when Achilles arrives at $x_0$. By our set-up, this happens at the $t$-time 
\be
t_1=\frac{x_0}{v},
\ee
and by that time, the tortoise has moved to
\be
x_1 = x_0+w\cdot \frac{x_0}{v} = x_0\cdot\left[1+\frac{w}{v}\right].
\ee
Zeno's next snapshot, to which we will accordingly assign $T=2$, is when Achilles reaches $x_1$, which happens at
\be
t_2=\frac{x_1}{v}=\frac{x_0}{v}\cdot\left[1+\frac{w}{v}\right]
\ee
At this time, the tortoise has gone on to
\be
x_2 = x_0\cdot\left[1+\frac{w}{v}+\left(\frac{w}{v}\right)^2\right].
\ee
Continuing in the same manner, we find that $x_i$ is given by the first $i$ terms of the geometric series, so that 
\be
x_i=x_0\sum_{j=0}^{i}\left(\frac{w}{v}\right)^j= x_0 \left(\frac{1-(w/v)^{i+1}}{1-(w/v)}\right),
\label{eq:xiGeometric}
\ee
making use of the well-known close-form version of that partial series. Given that the time at which $x_{i-1}$ is reached by Achilles is $T=i$ and $t_i=x_{i-1}/v$, we can use (\ref{eq:xiGeometric}) to write down the relationship
\be
t(T) =\frac{x_0}{v} \left(\frac{1-(w/v)^{T}}{1-(w/v)}\right)
\label{eq:tfromT}
\ee
between the old time coordinate $t$ and the new one $T$. The relation (\ref{eq:tfromT}) is readily inverted to give the $x_0$-dependent family of Zeno time coordinates $T$
\be
T(t) = \frac{\ln\left[1-\left(\frac{vt}{x_0}\right)\left(1-w/v\right)\right]}{\ln(w/v)}.
\label{eq:Tfromt}
\ee
By its derivation, the relation and its inverse initially only holds at integer times $T$ and at specific $x$ locations. But clearly, the relation can be extended to all real values of $T$ and $t$. The time coordinate $t$ is already defined for all real values, and we can use (\ref{eq:Tfromt}) to extend the integer times $T$ to a continuous range. Adopting the same notion of simultaneity for $T$ as we had for $t$, we can view $T$ as an alternative time coordinate. These, if you will, are the original tortoise coordinates, sharing with the relativistic Regge-Wheeler version the characteristic logarithm.

\begin{figure}
\begin{center}
\includegraphics[width=0.6\linewidth]{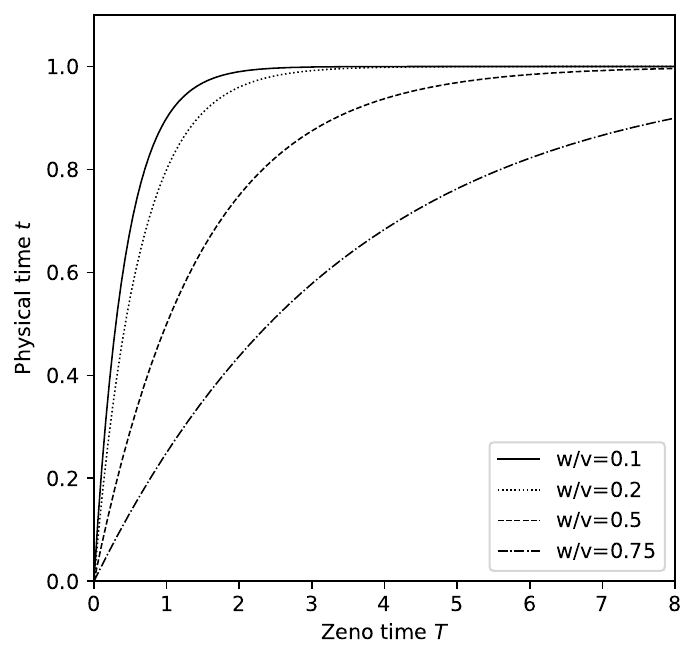}
\caption{Plotting physical time against Zeno time. For convenience, physical time units have been chosen so that $x_C/v=1$. \label{fig:FallPlot1}}
\end{center}
\end{figure}

\begin{figure}
\begin{center}
\includegraphics[width=0.6\linewidth]{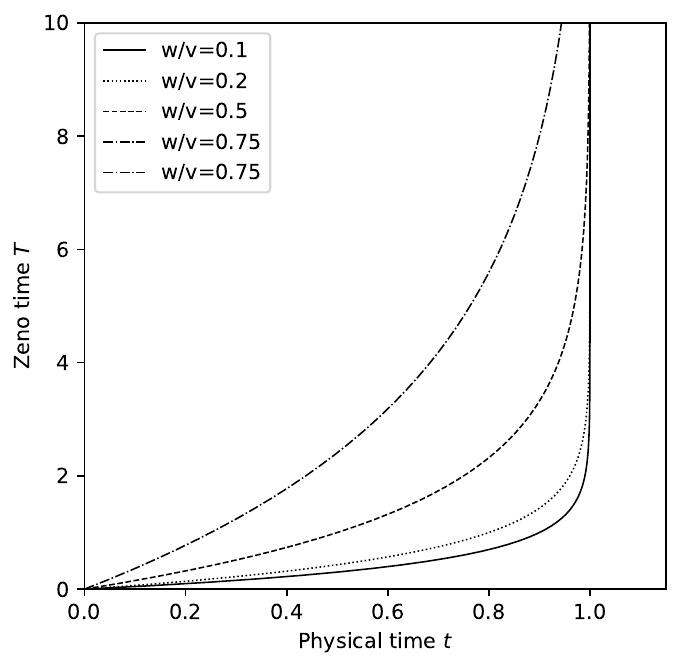}
\caption{Plotting Zeno time against physical time. Conventions as in Fig.~\ref{fig:FallPlot1} \label{fig:FallPlot2}}
\end{center}
\end{figure}

As Zeno's problem is formulated, it is natural to parametrise the situation in terms of the tortoise's initial head start $x_0$. But we can readily use a different parameter, namely the location where, weird coordinates apart, Achilles {\em does} catch up with the tortoise. This has the advantage of bringing us even closer to the black-hole situation, where it is the horizon location that we use to parametrize where strange things happen.

The time $t_C$ it takes Achilles to catch up is
\be
t_C = \frac{x_0}{v-w},
\ee
corresponding to the catch-up location
\be
x_C=vt_C =  \frac{x_0}{1-w/v}.
\ee
In terms of $x_C$, (\ref{eq:Tfromt}) simplifies to
\be
T(t) = \frac{\ln\left[1-\left(\frac{vt}{x_C}\right)\right]}{\ln(w/v)}.
\label{eq:Tfromt2}
\ee
This tells us immediately Zeno's version of the story: as Achilles approaches the ``tortoise horizon,'' $vt\to x_C$, we have $T\to\infty$. Plotting (\ref{eq:Tfromt2}) and its inverse can serve to drive the point home in a visual manner: In Fig.~\ref{fig:FallPlot1}, it becomes obvious how $t$ changes ever more slowly as we move to larger values of $T$, while in Fig.~\ref{fig:FallPlot2} it is clear that $T$ grows beyonds all bounds as we approach the horizon at reaching-the-tortoise-time.

This example and its simple one-dimensional coordinate transformation provide a suitable toy model that allows students to understand the limitations of Schwarzschild coordinates near a black hole, even without understanding the details of Schwarzschild coordinates. Zeno's story has the advantage that students intuitively know the right answer --- there is no doubt that Achilles {\em does} catch the tortoise. Seeing how an unusual time coordinate, which evidently covers only a limited region of the time axis, can lead to the (coordinate) statement that Achilles can never catch up prepares the students for understanding the Schwarzschild counterpart: that going by Schwarzschild coordinates, no object will ever reach the event horizon, but that this does not keep objects from falling in. 


\section*{Acknowledgements}

I would like to thank Thomas M\"uller for helpful comments on an earlier version of this text.\\

\end{document}